\documentclass[aps,twocolumn,pre,showpacs,superscriptaddress,groupedaddress,amsmath,amssymb]{revtex4-1}
\usepackage{epsfig}
\usepackage{graphics}
\usepackage{lipsum}
\usepackage{sidecap}
\usepackage{graphicx} %Include figure files
\usepackage{dcolumn} %Align table columns on decimal point
\usepackage{bm}% bold math
\usepackage{xcolor,colortbl}
\usepackage{color, colortbl}
\definecolor{darkgray}{rgb}{0.66, 0.66, 0.66}
\definecolor{yellow-green}{rgb}{0.6, 0.8, 0.2}
\definecolor{deeppink}{rgb}{1.0, 0.08, 0.58}
\definecolor{darkviolet}{rgb}{0.58, 0.0, 0.83}
\definecolor{darkcyan}{rgb}{0.0, 0.55, 0.55}

\usepackage{sidecap}  
\begin{document}
\preprint{APS/123-QED}

\title{Effect of asymmetry parameter on the dynamical states of nonlocally coupled nonlinear oscillators}

\author{R.~Gopal$^{1,2}$}
\author{V.~K.~Chandrasekar$^{3}$}
\author{D.~V.~Senthilkumar$^{3}$}
\author{A.~Venkatesan$^{2}$}
\author{M.~Lakshmanan$^1$}
 
\affiliation{
$^{1}$Centre for Nonlinear Dynamics, School of Physics, Bharathidasan University, Tiruchirapalli-620024, India\\
$^{2}$Department of Physics, Nehru Memorial College, Puthanampatti,
Tiruchirapalli 621 007, India.\\
$^{3}$Centre for Nonlinear Science \& Engineering, School of Electrical \& Electronics Engineering, SASTRA University, Thanjavur- 613 401, India.
}
\date{\today}
            
\begin{abstract}
We show that  coexisting domains of coherent and incoherent oscillations can be induced in an
ensemble of any identical nonlinear dynamical systems using the nonlocal rotational matrix coupling
with an asymmetry parameter.
Further, chimera is shown to emerge in a wide range of the asymmetry parameter in contrast to
near $\frac{\pi}{2}$ values of it employed in the earlier works.
We have also corroborated our results using the strength of incoherence in the frequency domain ($S_{\omega}$)
and in the amplitude domain ($S$) thereby distinguishing the frequency and amplitude chimeras.
The robust nature of the asymmetry parameter in inducing chimeras in any generic dynamical system
is established using  ensembles of  identical R\"ossler oscillators, Lorenz systems, and
Hindmarsh-Rose (HR) neurons in their chaotic regimes.
\end{abstract}


\maketitle

\section{Introduction}

A fascinating emergent phenomenon in a network of nonlocally coupled identical
oscillators is characterized by the simultaneous existence of synchronous and desynchronous domains of
oscillators ~\cite{kuramoto2002}. This phenomenon was first observed by Kuramoto and 
Battogtokh in a model of complex Ginzburg-Landau equation with exponentially decaying 
nonlocal coupling~\cite{idbyk2000,kuramoto2002,free1967,kuramoto1984}. 
Later investigations by
Abrams and Strogatz revealed similar states in an ensemble of phase oscillators with nonlocal 
coupling, who coined the term ``chimera'' for such states with coexisting 
coherent and incoherent domains~\cite{abrams2004}. Since then, identifying chimera states and
 their existence criteria has become an active area of research both theoretically and experimentally
~\cite{abrams2004,abrams2006}.

In particular, chimera states were observed in networks of identical, symmetrically 
coupled Kuramoto phase oscillators, coupled Stuart-Landau oscillators with nonlocal 
interactions in one and two-dimensional arrays and also in systems  with 
delay coupling~\cite{abrams2008,sethia2008,sethia2013}.  Recently, it has been 
shown that a symmetry breaking coupling in Stuart-Landau oscillators leads to 
the manifestation of chimera death~\cite{zak2014}. Very recently, the existence 
of chimera states have also been demonstrated experimentally in populations of coupled  
chemical oscillators, in electro-optical coupled map lattices by liquid-crystal 
light modulators~\cite{tinsley2012,hagers2012} and in a mechanical experiment involving 
two sub-populations of identical metronomes coupled in a hierarchical 
network~\cite{martens2013}. Real world examples exhibiting the states mimicking chimera states
include the unihemispheric sleep of animals~\cite{rottenberg2000}, 
power grids~\cite{Filatrella2008}, and so on.  Understanding intricacies involved in 
the dynamics of chimera states 
is also very important from the perspective of neuroscience, as it is believed that it can
be associated with the concept of ``bumps" of neuronal activity~\cite{laing2000}.

Initial investigations on the phenomenon of chimera have adopted an ensemble of phase-only models 
in the weak coupling limit and demonstrated  that chimera will arise when the ensemble is distributed
spatially using  a nonlocal coupling with a well tuned Sakaguchi phase-lag parameter, which
introduces a phase asymmetry~\cite{abrams2004,abrams2008,gcsas2008}. 
In particular, it was established that chimera states can arise 
typically when the phase-lag parameter is near $\frac{\pi}{2}$. Later investigations have extended the
notion of chimera beyond the phase-only models both in the case of nonlocal coupling as well as global coupling ~\cite{ioym2011,rgvkc2014,azmk2014}. Interestingly,
a couple of recent investigations~\cite{omel2013,avjh2014} have considered dynamical systems that can be reduced to
phase models and employed a nonlocal rotational matrix coupling with a phase-lag parameter, similar to the Sakaguchi phase-lag parameter. Correspondingly, the dynamical systems in their periodic regimes exhibit transitions, for appropriate value (near $\frac{\pi}{2}$) of
the phase-lag parameter, observed in the phase-only models including chimera states. 
However, it is not clear whether the nonlocal rotational matrix coupling can indeed
induce chimera in a generic dynamical system that may not be reduced to phase models
or even in dynamical systems that can be reduced to phase models but in the aperiodic
(chaotic) regimes.

\begin{figure*}
\centering
\includegraphics[width=2.1\columnwidth]{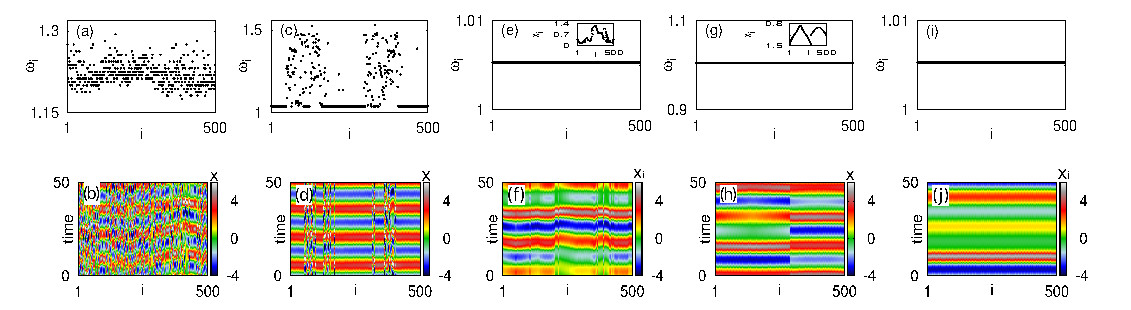}

\caption{(Color online)~Time averaged frequency of the oscillators (top row) and  spatiotemporal 
plots (bottom row) as a function of the oscillator index $i=1,\ldots,N=500$ for
an ensemble of R\"ossler oscillators with nonlocal rotational matrix coupling for
coupling radius $r=0.3$, coupling strength $\varepsilon=0.1$ and for 
different values of the asymmetry parameter $\alpha$.
(a) \& (b) completely asynchronous oscillations for $\alpha=1.57$,  
(c) \& (d) coexisting domains of coherent and incoherent oscillations illustrating
frequency chimera for $\alpha=1.47$, 
(e) \& (f) frequencies are synchronized while there exists a drift in the 
amplitude of the oscillations, which are clearly indicated by domains of coherent and
incoherent oscillations in the snap shot depicted in the inset, for $\alpha=0.97$ corresponding
to amplitude chimera,
(g) \& (h) cluster states with frequency entrainment for $\alpha=0.57$, and 
(i) \& (j) the ensemble of oscillators are completely synchronized for $\alpha=0.1$.}
\label{fig1}
\end{figure*}
Motivated by the above, in this paper we consider the collective dynamics of ensembles of identical nonlinear oscillators interacting via nonlocal rotational matrix coupling. For vanishing value of the asymmetry (phase-lag) parameter in the rotational matrix coupling, the
ensemble of dynamical systems is characterized by a direct coupling and it exhibits a global
synchronization above a threshold value of the coupling strength. In contrast, for 
$\frac{\pi}{2}$, the ensemble is driven by a cross (conjugate) coupling rendering 
the ensemble to exhibit  asynchronous oscillations for any value of the coupling strength. 
For intermediate values of the asymmetry parameter, we find a  wide range of
collective dynamical behaviors including two different types of chimeras, thereby 
elucidating the emergence of chimeras in a wide range of the asymmetry parameter 
in contrast to the earlier studies sticking to near $\frac{\pi}{2}$ values of  
the asymmetry parameter. In particular, when the asymmetry parameter is increased 
from a zero value, for appropriate coupling strengths, we find the occurrence of 
both frequency and amplitude chimeras in generic dynamical systems that cannot be 
reduced to phase models. In addition, a gallery of collective  
dynamical states including clusters, coherence and fully synchronous oscillations
arise in these systems. On the other hand, in the earlier investigations 
frequency chimera alone is shown to exist in the periodic 
regimes with nonlocal rotational matrix coupling in systems that can be
reduced to phase models~\cite {omel2013}. Further, the chimera states
are characterized by inhomogeneous spread of coherent and incoherent domains
unlike homogeneous spacing of coherent and incoherent domains in phase models
and in periodic oscillators. 
 We establish these facts using three paradigmatic models, namely coupled R\"ossler oscillators, Lorenz systems, and Hindmarsh-Rose (HR) neurons with nonlocal and rotational matrix coupling.

The plan of the article is as follows. An ensemble of identical R\"ossler oscillators
will be introduced in Sec. II, where the role of the asymmetry parameter in the nonlocal
rotational matrix coupling in inducing frequency and amplitude chimeras will be discussed.
Section III deals with the quantitative characterization by introducing the strength of incoherence
in frequency and amplitude domains thereby distinguishing frequency and amplitude chimeras,
respectively. Global bifurcation scenario will be discussed  in Sec. IV and  the generic
nature of the asymmetry parameter in inducing chimeras in a wide range of it will be
established using  ensembles of Lorenz oscillators and Hindmarsh-Rose neurons in Sec. V.
Finally a summary of the results and discussions will be provided in the concluding section VI.

\begin{figure}
\centering
\includegraphics[width=0.9\columnwidth]{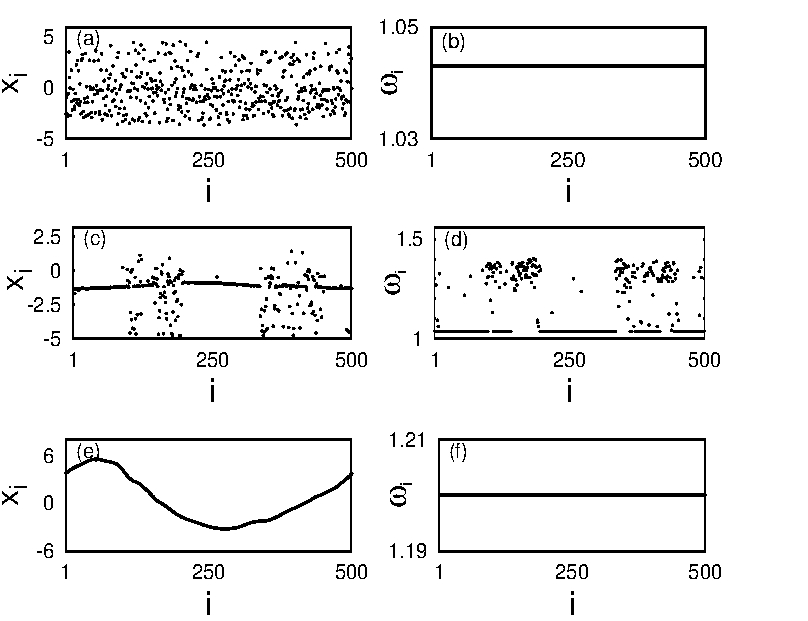}
\caption{Snapshots of the variables $x_{i}$ (left column) and the time averaged frequency of 
the oscillators (right column) of the ensemble of R\"ossler oscillators
for the coupling radius $r=0.3$, the asymmetry parameter $\alpha=1.47$ and
for different values of the coupling strength $\varepsilon$. 
(a)  incoherent amplitude for $\varepsilon=0.05$, 
(b)  coherent frequency for $\varepsilon=0.05$,
(c) coexisting coherent and incoherent domains of amplitude for $\varepsilon=0.22$,
 (d) coexisting coherent and incoherent domains of frequency for $\varepsilon=0.22$, 
(e)   coherent oscillations with fixed phase difference with the nearby oscillators for $\varepsilon=0.35$
and (f) coherent frequency for $\varepsilon=0.35$.}
\label{fig2}
\end{figure}
\section{Frequency and amplitude chimera in an ensemble of R\"ossler oscillators}
We consider the following ensemble of R\"ossler oscillators with nonlocal, and rotational matrix coupling 
\begin{subequations}
\begin{align}
\dot{x}_{i}=&\,-y_{i}-z_{i}+\frac{\epsilon}{2P}\sum_{j=i-P}^{j=i+P}~[\alpha_{11}(x_{j}-x_{i})+\alpha_{12}(y_{j}-y_{i})]\\
\dot{y}_{i}=&\,x_{i}+ay_{i}+\frac{\epsilon}{2P}\sum_{j=i-P}^{j=i+P}~[\alpha_{21}(x_{j}-x_{i})+\alpha_{22}(y_{j}-y_{i})] \\
\dot{z}_{i}=&\,b+(x_{i}-c)z_{i} 
\end{align}
\label{eq1}
\end{subequations}
where $i=1,2,...,N=500$. The system parameters are chosen as $a=0.42$, $b=2$ and $c=4$, 
for which the individual R\"ossler oscillators exhibit chaotic oscillations. $\varepsilon$ 
is the coupling strength  and $P\in(1,N/2)$ is the number of nearest neighbors on each 
side of any oscillator in the ring with a coupling radius $r=\frac{P}{N}$, which provides the measure of nonlocal coupling.  The coefficients $\alpha_{lk}$, where $l,k \in 1,2$, are the 
components of  the rotational matrix coupling represented as
\begin{equation}
{\bf{B}}=\left( \begin{array}{cc}
\alpha_{11} & \alpha_{12}  \\
\alpha_{21} & \alpha_{22} \\
 \end{array} \right) = \left( \begin{array}{cc}
cos\alpha & sin\alpha  \\
-sin\alpha & cos\alpha \\
 \end{array} \right).
\end{equation}
Here $\alpha$ is the asymmetry parameter. The matrix ${\bf{B}}$ facilitates both the
direct coupling as well as the cross coupling between the variables $x$ and $y$. 
For $\alpha=\frac{\pi}{2}$, one has the cross (conjugate) coupling where the $y$ 
variable is coupled to $x$ and $x$ to $y$ and the ensemble is in complete asynchrony 
(see Figs.~\ref{fig1}(a) and (b)) in the entire range of $\varepsilon$.  
On the otherhand, the coupling is direct for $\alpha=0$ with the $x$ 
variable coupled to $x$ and $y$ to $y$, for which the ensemble of coupled 
systems, Eq.~(\ref{eq1}), is in complete synchrony for $\varepsilon>0.1$ 
(see Figs.~\ref{fig1}(i) and (j)), while for intermediate values of $\alpha$,
different other interesting dynamical states occur (see Figs.~\ref{fig1}(c) and (h)) 
as described below. Thus the asymmetry parameter $\alpha$ plays a crucial role in
determining  the collective dynamics of the ensemble of coupled systems with nonlocal rotational matrix
coupling. For intermediate values of the asymmetry parameter  $\alpha\in[0,\frac{\pi}{2}]$, the 
coupling in Eq.~(\ref{eq1}) includes both direct and cross couplings with the value 
of the weights  $\alpha_{lk}$ determining the proportion of them  thereby dictating the collective dynamics of the ensemble.  For suitable intermediate values of the asymmetry parameter, the ensemble exhibits a mixture of the above extremities thereby leading to the emergence of coexisting coherent 
and incoherent domains by a spontaneous splitting of the ensemble. 
In particular, coherence and incoherence can be found
both in the amplitude and frequency, and  also incoherence in the amplitude alone but with coherent frequency,
thereby distinguishing frequency chimera and amplitude chimera, respectively. 
Considering chaotically evolving coupled oscillators, the frequency chimera is characterized by both temporal and spatial chaos in the
incoherent interval whereas only spatial chaos is observed in the incoherent interval of
the amplitude chimera while the corresponding temporal dynamics being periodic in most cases~\cite{Dud2014}. Beside the above two types of chimeras, cluster states, coherent states and completely synchronized states are also
observed during the dynamical transitions as discussed below.

\begin{figure}
\centering
\includegraphics[width=0.9\columnwidth]{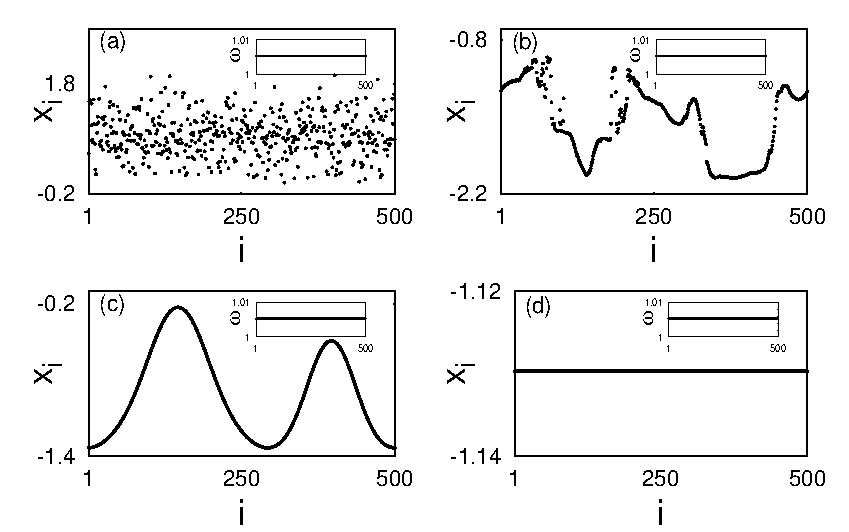}
\caption{Same as Fig.~\ref{fig2} but for the value of the asymmetry parameter $\alpha=0.95$.
In this case, the time averaged frequencies of the oscillators are in coherence,
for all values of the coupling strength we have examined, as shown in the insets
of the figures. (a) asynchronous oscillations of the ensemble of R\"ossler oscillators for
$\varepsilon=0.03$, (b) coexisting domains of synchronous and asynchronous oscillations for
$\varepsilon=0.1$, (c) coherent oscillations for $\varepsilon=0.2$, and (d) 
complete synchronous oscillations for $\varepsilon=0.3$.}
\label{fig3}
\end{figure}

We have fixed the coupling radius as $r=0.3$ to investigate the dynamical transitions
by gradually decreasing the asymmetry parameter from $\alpha=\frac{\pi}{2}=1.57$ 
for the value of coupling strength $\varepsilon=0.1$. The ensemble of identical 
oscillators exhibits a complete asynchrony for $\alpha=\frac{\pi}{2}$ as discussed above
for any value of the coupling strength. The cross (conjugate) coupling induces a
drift in the oscillations of identical oscillators resulting in the distribution of
their frequencies in a large range as shown in Figs.~\ref{fig1}(a) and (b). 
 The values of $\omega_i$ for each oscillator is calculated as 
$\omega_i=2\pi K_i/\Delta t, i=1, 2, 3, \ldots, N$, where $K_i$ is the number of maxima
of the time series $x_i(t)$ of the $i^{th}$ oscillator during the time interval $\Delta t$.
Decreasing
the asymmetry parameter to $\alpha=1.47$, a small fraction of the direct coupling competes
with the conjugate coupling in reinforcing certain oscillators to oscillate in synchrony
resulting in a group of synchronized oscillators while retaining the rest in asynchrony (see Figs.~\ref{fig1}(c) and (d)).
This scenario of spontaneous splitting of the ensemble into coexisting coherent and
incoherent domains is nothing but the chimera state. It is evident from Figs.~\ref{fig1}(c) and (d)
that both the frequency and the amplitude of all the oscillators are in synchrony in the coherent domain,
while in the other domain they remain desynchronized thereby characterizing the  underlying state as a  frequency chimera.
It is also clear from these figures that there is an inhomogenity in
the spread of the coherent and incoherent domains unlike the homogeneous spread of both the domains
in phase only models and in the case of periodic oscillators.
Upon decreasing $\alpha$ further to $\alpha=0.97$, the proportion of the direct and the conjugate 
couplings are close to each other and in this region the ensemble
is entrained to a single frequency resulting in complete coherence (see Fig.~\ref{fig1}(e)). 
Nevertheless, the amplitudes of all the oscillators are not completely locked resulting in 
partially coherent and incoherent domains in amplitude as depicted in the inset of Fig.~\ref{fig1}(e),
which is also clearly visualized in the corresponding spatiotemporal plot in  Fig.~\ref{fig1}(f).
This state with the coexistence of oscillators with coherent and incoherent amplitudes while their
frequencies are completely entrained is classified as an amplitude chimera.  
Synchronized clusters emerge for further decrease in the asymmetry parameter 
as illustrated in  Figs.~\ref{fig1}(g) and (h) for $\alpha=0.57$, where the fraction of the
direct coupling is higher than that of the conjugate coupling. When the direct
coupling dominates the conjugate coupling for further lower values of the asymmetry parameter 
the ensemble achieves complete synchronization, which is shown in Figs.~\ref{fig1}(i) and (j)
for $\alpha=0.1$.  

\begin{figure*}
\centering
\includegraphics[width=1.5\columnwidth]{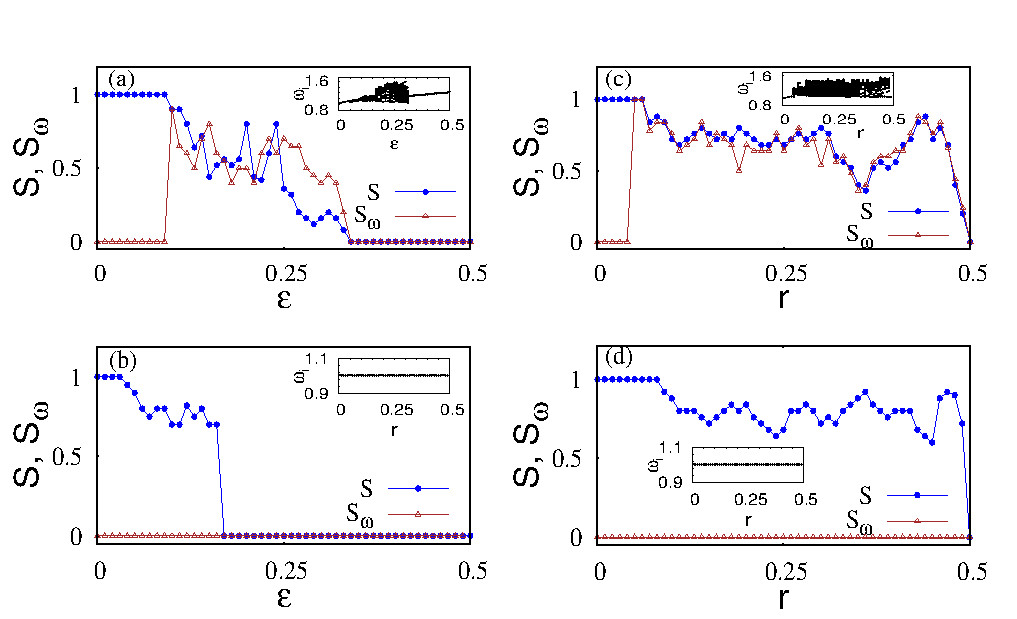}
\caption{(Color online)~(a) and (b) Strength of incoherence in amplitude $S$  and in frequency $S_{\omega}$ as a function
of the coupling strength characterizing the dynamical transitions discussed in
Figs.~\ref{fig2} and \ref{fig3}, respectively, for $r=0.3$.
Time averaged frequency of the oscillators are shown in  the insets. In (c) and (d), $S$  and $S_{\omega}$ are plotted as a function of
the coupling radius $r$ for the fixed coupling strengths $\varepsilon=0.2$ and $0.1$, respectively. The value of the asymmetry parameter has been fixed as $\alpha=1.47$ in (a) and (c), while it is chosen as $\alpha=0.95$ in (b) and (d).
Degree of disorder in the oscillator frequencies is increased in the range of frequency chimera as
evident from the insets in (a) and (c).}
\label{fig4}
\end{figure*}
Now we will trace the dynamical transitions by varying the coupling strength for a couple
of fixed values of the asymmetry parameter $\alpha$. Snapshots and average frequencies of
the ensemble of oscillators are shown in the left and right column of Fig.~\ref{fig2}, respectively, as 
a function of the oscillator index for $\alpha=1.47$ for different values of the coupling strength.
The amplitude of the oscillators are completely uncorrelated for $\varepsilon=0.05$ as can be seen
in Fig.~\ref{fig2}(a), even while their frequencies are entrained (see Fig.~\ref{fig2}(b)).
Upon increasing the coupling strength further, frequency chimera comes into existence as
characterized by coexisting asynchronous and synchronous domains in both the amplitude (Fig.~\ref{fig2}(c))
and the frequency (Fig.~\ref{fig2}(d)) of the oscillators. Finally, all the oscillators
evolve in coherence as depicted in Fig.~\ref{fig2}(e) with synchronized frequency (see Fig.~\ref{fig2}(f)).
Next, we will fix the asymmetry parameter at $\alpha=0.95$, where we observe the amplitude chimera.
Snapshots along with the frequency profile of all the oscillators are depicted in  Figs.~\ref{fig3}.
The frequencies of
the oscillator ensemble are always entrained  for the chosen value of the asymmetry parameter as
can be seen from the insets of Figs.~\ref{fig3}. The oscillators evolve independently
for $\varepsilon=0.03$, which is evident from the random values acquired by the oscillator ensemble
depicted in Fig.~\ref{fig3}(a). Amplitude chimera emerges for  $\varepsilon=0.1$ (see Fig.~\ref{fig3}(b)), 
where the oscillators are grouped into synchronous and asynchronous domains depending upon the amplitude correlation.
Eventually, all the oscillators evolve in coherence with fixed phase difference with the nearby oscillators
for further higher values of the coupling strength, the snapshot of which is shown 
in Fig.~\ref{fig3}(c) for  $\varepsilon=0.2$. Finally all the oscillators are in complete synchronization for $\varepsilon>0.23$ as depicted in Fig.~\ref{fig3}(d) for $\varepsilon=0.3$.

\begin{figure}
\centering
\includegraphics[width=0.8\columnwidth]{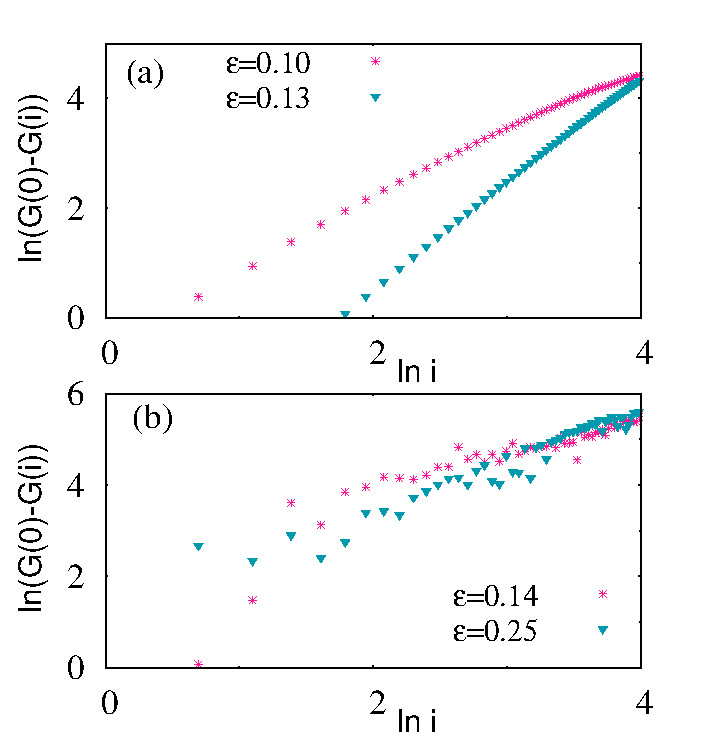}
\caption{(Color online)~~Plots of the spatial correlation function $G(0)-G(i)$ (a) for amplitude chimera for $r=0.3$ and $\alpha=0.95$, showing a power-law dependence, and (b) for frequency chimera for $r=0.3$ and $\alpha=1.47$, showing no power-law dependence.}
\label{fig4a}
\end{figure}

%%%

\section{Quantitative characterization of chimera states in the ensemble of R\"ossler oscillators}
Recently, we have introduced the strength of incoherence $S$~\cite{rgvkc2014}, 
as a measure to distinguish various collective dynamical states, which is defined as
\begin{equation} 
S=1-\frac{\sum_{m=1}^{M}s_{m}}{M}, \hspace{0.1cm}  s_{m}=\Theta(\delta-\sigma_{l}(m)),
\end{equation}
where $\Theta(\cdot)$ is the Heaviside step function, and $\delta$ is a predefined 
threshold. Here, we take $\delta$ as a certain percentage value of difference between
the upper/lower bounds, $x_{l,i,max/min}$,  of the allowed values of $x_{l,i}$'s.
 $M=20$ is the number of bins of equal size $n=N/M=25$, where $N$ is the total number of
oscillators.
The local standard deviation $\sigma_{l}(m)$ is introduced as
\begin{equation}
\sigma_{l}(m)=\Big<\noindent \sqrt{\frac{1}{n}\sum_{j=n(m-1)+1}^{mn}[z_{l,j}-<z_{l,m}>]^2} 
\hspace{0.1cm}\Big>_{t}, \hspace{0.1cm}m=1,2,...M,
\end{equation}
where $z_{l,i}=x_{l,i}-x_{l,i+1}$, $l=1,2...d$, $d$ is the dimension of the individual unit in
the ensemble, $i=1,2...N$, $<z_{l,m}>=\frac{1}{n}\sum_{j=n(m-1)+1}^{mn}z_{l,j}(t)$, and
$\langle ...\rangle_{t}$ denotes average over time.
When $\sigma_{l}(m)$ is less than $\delta$, $s_{m}=1$, otherwise it is `$0$'. 
 In the incoherent domain, the local standard deviation $\sigma_{l}(m)$ has 
some finite value $\forall m$, which is always greater than the predefined threshold
$\delta$ and hence  $s_{m}=0, \forall m$, thereby resulting in unit value for the
strength of incoherence $S$ in this domain. On the other hand, in the case of coherent domain the
standard deviation $\sigma_{l}(m)$ is always zero in the coherent domain and 
hence $s_{m}=1, \forall m$, thereby resulting in a zero value for $S$ for this domain.
Since the chimera states are characterized by coexisting coherent and incoherent domains,
the strength of incoherence $S$ will have intermediate values 
between zero and one, $0 < S < 1$. However, the
strength of incoherence is incapable of distinguishing different types of chimera.
In order to facilitate this, we estimate the strength of incoherence in the frequency domain
$S_{\omega}$, as different from $S$ where we use the same prescriptions as above with the choice $z_{l,i}=\omega_{l,i}-\omega_{l,i+1}$.  Now, $S_{\omega}$
can be used to clearly distinguish amplitude and frequency chimera. Since the frequencies
of all the oscillators in the ensemble is entrained for amplitude chimera, the strength of incoherence in the frequency domain for the amplitude chimera is always $S_{\omega}=0$, whereas it varies between $0 < S_{\omega} < 1$ for frequency
chimera.

$S$ and $S_{\omega}$ are depicted in Figs.~\ref{fig4}(a) and (b) as a function of
the coupling strength $\varepsilon$ for two different values of the asymmetry parameter
$\alpha=1.47$ and $0.95$, respectively, characterizing the dynamical transitions
discussed in Figs.~\ref{fig2} and~\ref{fig3}  for the specific coupling radius $r=0.3$.  In the range $\varepsilon\in(0,0.09)$
the unit value of $S$ indicates that the oscillators evolve independently while at the same time
$S_{\omega}=0$ corroborates that the frequency of the oscillators are synchronized. 
Fluctuation in the values of $S$ and $S_{\omega}$ between null value and unity 
confirms the existence of frequency chimera in the range of $\varepsilon\in(0.09,0.33)$. Beyond
$\varepsilon=0.33$ both $S$ and $S_{\omega}$ attain null value corresponding to
the existence of complete coherence among the oscillators. It is to be
noted that the frequency profile of the oscillators is distributed over a wide range
in the frequency chimera state thereby showing an increase in the degree of disorder among the
oscillator frequencies (see the inset of Fig.~\ref{fig4}(a)).  As discussed earlier
in Fig.~\ref{fig3} for $\alpha=0.95$, the frequency of the oscillator ensemble
is always entrained as indicated by the null value of $S_{\omega}$ in Fig.~\ref{fig4}(b),
whereas up to $\varepsilon=0.03$, $S$ acquires unit value attributing to asynchronous
amplitude variation. Amplitude chimera exists in the range  $\varepsilon\in(0.03,0.16)$,
where $S$ fluctuates between zero and unity while $S_{\omega}=0$ in this range. All the oscillators evolve in coherence
for  $\varepsilon>0.16$, which is confirmed by the null values of $S$ and $S_{\omega}$ in Fig.~\ref{fig4}(b).

Next, the strengths of incoherence $S$ and $S_{\omega}$ are shown in Figs.~\ref{fig4}(c) and (d) as a function of the coupling radius $r$ for two different values of the asymmetry parameter
$\alpha=1.47$ and $0.95$, respectively. 
The intermediate values of $S$ and $S_{\omega}$ between  zero  and one
in Fig.~\ref{fig4}(c) for $\varepsilon=0.2$ characterize the existence of frequency
chimera in a wide range of the coupling radius $r\in(0.07,0.49)$. It is to be noted
that for $r=0.5$, the nonlocal coupling turns out to be the global coupling. 
The value of $S_{\omega}$ in Fig.~\ref{fig4}(d) for $\varepsilon=0.1$ remains zero 
 while $S$ fluctuates between zero and unity in the range of $r\in(0.09,0.49)$,
elucidating the existence of amplitude chimera as a function of the coupling radius.
Thus, it is clearly evident from both these figures that chimera states can indeed
appear in a rather wide range of the coupling radius $r$.
Further, the frequency profile of the oscillators is distributed over a wide range
in the frequency chimera state thereby showing an increase in the degree of disorder among the
oscillator frequencies as shown in the inset of Fig.~\ref{fig4}(c).

\begin{figure}
\centering
\includegraphics[width=0.9\columnwidth]{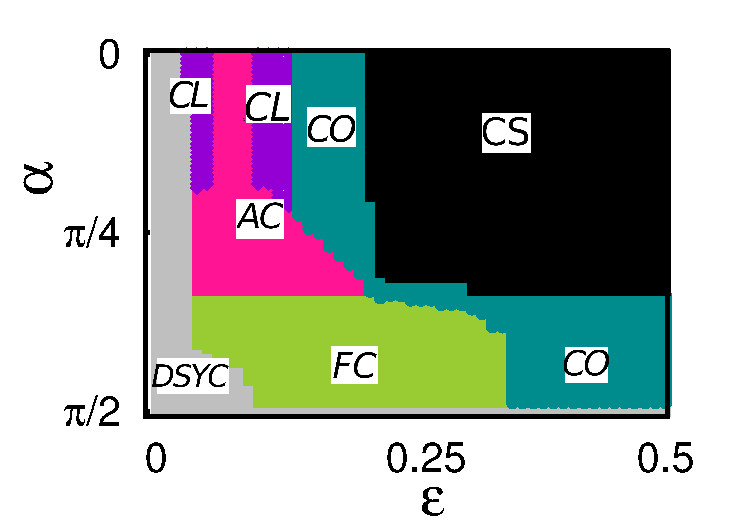}
\caption{(Color online)~Two parameter phase diagram for the dynamical states of the R\"ossler 
oscillators (\ref{eq1}) as a function of the coupling strength
$\epsilon$ and the asymmetry parameter $\alpha$. See Table I and 
text for details. Note that the power-law scaling turbulence exists only in the amplitude chimera region ($AC$) as explained in the text.}
\label{fig5}
\end{figure}
\begin{table}
\caption{(Online color) Characterization of different collective states with the corresponding color coding in Figs. 6,8 and 10:}
\begin{ruledtabular}
\begin{tabular}{llllll}
Dynamical state &~~~ $S$  &~~~ $S_{\omega}$  & ~~~$S_{0}$ &~~~ $S_{x}$ &~~~ Remarks \\
\hline
\color{gray}
{DSYC}          &~~~   1  &~~~    0         &~~~ 1  &~~~   1  &~~~     \\
\color{yellow-green}
 {FC}             &~~~   c  &~~~    c         &~~~  c  &~~~   d  &~~~  $0<c,d<1$          \\
\color{deeppink}
 {AC}            &~~~   c  &~~~    0         &~~~  c  &~~~     d  &~~~   \\
\color{darkviolet}
 {CL}             &~~~   c  &~~~    0         &~~~  0  &~~~    d  & ~~~         \\  
\color{darkcyan}
 {CO}           &~~~   0  & ~~~   0         &~~~  0  & ~~~      d &~~~ \\

 CS             & ~~~  0  & ~~~   0         & ~~~ 0  &~~~     0  &~~~      \\ 
\end{tabular}
\end{ruledtabular}
\end{table}

In order to check how the amplitude and frequency chimeras characterized by the 
strengths of incoherence  as discussed above relate to the power-law scaling 
turbulence in the nonlocally coupled chaotic systems studied by Kuramoto and 
Nakao in Ref.~\cite{kura1996},  we have estimated the spatial correlation 
function $G(i)=\left<\sum_{j=0}^N x_jx_{j+i}\right>_t, i=0,1,...\frac{N}{2}-1, x_{j+N}=x_j$,
where $N$ is the total number of oscillators in the ensemble.
It has been demonstrated in~\cite{kura1996} that the power-law spatial correlations
appear generically in self-oscillatory media with nonlocal coupling exhibiting spatiotemporal chaos.
In particular, for small but finite $i$, the correlation function $G(i)$ behaves
like $G(0)-G(i)\approx \gamma_0+\gamma_1i^{\xi}, 0\le\xi\le2$, where $\gamma_0$ 
and  $\gamma_1$ are constants, exhibiting a power-law 
dependence on the distance and a discontinuous peak ($\gamma_0\ne0$) at the origin when the coupling
constant decreases below a critical value. We have found that the spatial correlation
function  $G(0)-G(i)$ exhibits a power-law  dependence on the distance $i$ as shown in Fig.~\ref{fig4a}(a) for the case of
amplitude chimera characterizing spatial chaos, whereas it does not have a power-law distribution in the case of frequency chimera (see Fig.~\ref{fig4a}(b)). We have also checked that spatial correlation function indeed
exhibits a power-law distribution in the entire parameter space of the amplitude chimera corresponding to Fig. 4(b).
 Nevertheless, we have confirmed that both the amplitude and frequency chimeras are characterized by a
discontinuous peak at the origin corroborating the existence of spatial incoherence as demonstrated
by Kuramoto and Nakao.

\section{Global bifurcation diagram for the R\"ossler oscillators}
The global scenario including the discussed dynamical transitions as a function 
of the coupling strength in the range  $\varepsilon\in(0,0.5)$ and the asymmetry parameter 
$\alpha\in(0,\frac{\pi}{2})$ is depicted in  Fig.~\ref{fig5}. The parameter regions for
which the oscillators are in the desynchronized state, cluster state, coherent state, and complete synchronization are marked as `$DSYC$',  `$CL$' `$CO$' and `$CS$', respectively, while that for the amplitude and  frequency chimeras they are indicated  by `$AC$' and `$FC$', respectively.  The corresponding color codeing and the nature of various strengths of incoherence are explained in Table I.

 In Fig.\ref{fig5}, the parameter space is demarcated using the strength of incoherence  $S$ to distinguish the incoherent, coherent and amplitude chimera regimes.  $S_{\omega}$ is used to distinguish
the frequency chimera from the amplitude chimera.  Method of removal of discontinuity ~\cite{rgvkc2014}
is used to distinguish cluster states from the other states. In this case, we introduce another measure $S_{0}$ which is the strength of incoherence after removal of discontinuities, while the previous measure $S$ corresponds to the strength of incoherence before the removal of discontinuity. The values of $S$ and $S_{0}$ indicates the existence of cluster state.
 The strength of incoherence $S_{x}$ obtained by using the values of the dynamical variables $x_{l,i}$ directly instead of the difference variables $z_{l,i}=x_{l,i}-x_{l,i+1}$ distinguishes between complete synchronization $(CS)$ with $S_{x}=0$, $S=0$ and coherent state $(CO)$ with $S_{x}=d$, $S=0$, $0<d<1$. We have used the same set of initial conditions, which is uniformly distributed between $1$ and $-1$, for all combinations of the parameters in the entire parameter space.

For $\alpha=\frac{\pi}{2}$, the ensemble of R\"ossler oscillators evolves asynchronously for the entire range of the coupling strength. In the range of the asymmetry parameter $\alpha=(1.55,0.97)$
transition from desynchronized state to frequency chimera and finally to coherent state
is observed as the coupling strength is varied in the range $\varepsilon\in(0,0.5)$. 
Decreasing $\alpha$ further, thereby increasing the proportion of direct coupling 
in the range $\alpha=(0.97,0.6)$, the ensemble of oscillators transit
to amplitude chimera from asynchronous state as the coupling strength is increased
from zero.  Further increase in the value of $\varepsilon$ leads to complete
coherence among the oscillators and finally to complete synchronous oscillations.
For $0<\alpha<0.6$, we observe similar dynamical transitions except for the fact that the 
amplitude chimera is surrounded by the cluster states in  both the directions of the
coupling strength as seen in  Fig.~\ref{fig5}. 

\begin{figure}
\centering
\includegraphics[width=0.9\columnwidth]{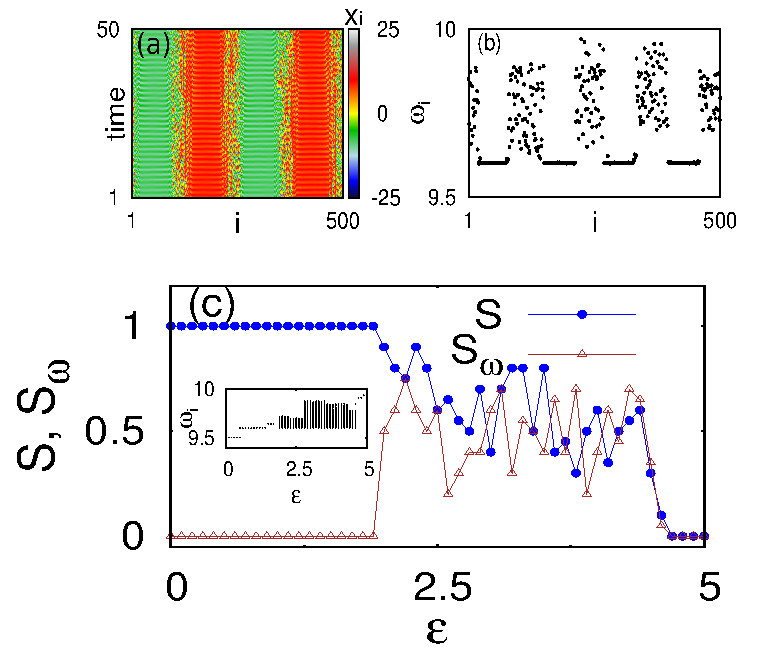}
\caption{(Color online)~(a) Space time plot and (b) time averaged frequency as a function of
the oscillator index for $\varepsilon=3.5$, and (c) $S$ and  $S_{\omega}$ as a function of the
coupling strength elucidating the frequency chimera in an ensemble of identical
Lorenz oscillators with nonlocal rotational matrix coupling  for $r=0.35$ and $\alpha=1.46$.}
\label{fig6}
\end{figure}
\begin{figure}
\centering
\includegraphics[width=0.9\columnwidth]{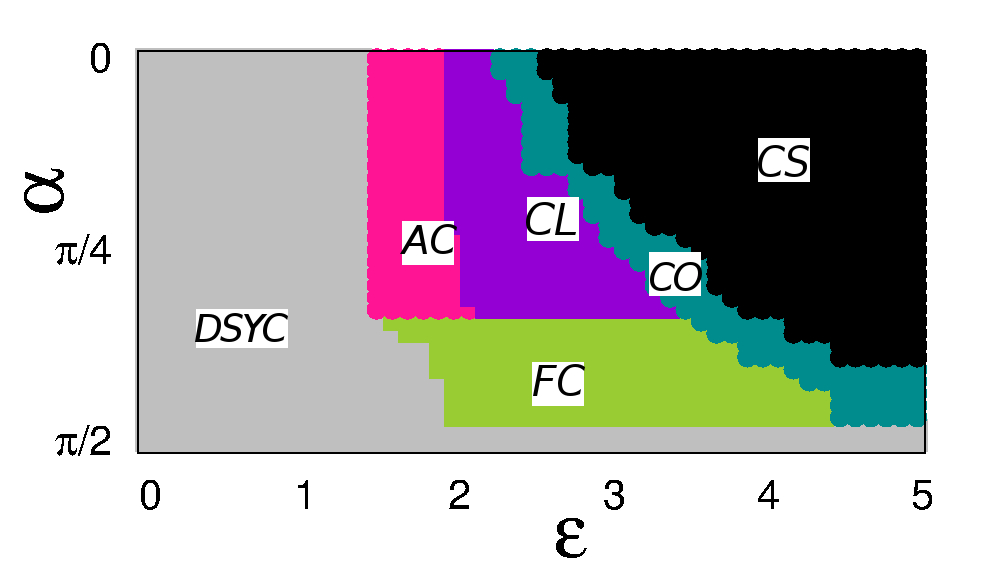}
\caption{(Color online)~Two parameter phase diagram for the dynamical states of the Lorenz
oscillators (\ref{eqlor}) as a function of the coupling strength
$\epsilon$ and the asymmetry parameter $\alpha$. See Table I and 
text for details.}
\label{fig6a}
\end{figure}
\section{Chimera in ensembles of Lorenz oscillators and Hindmarsh-Rose neurons}
In this section, we will establish our claim that the rotational matrix coupling 
with the asymmetry parameter $\alpha$ is capable of inducing chimera in any dynamical
system for  a wide range of values of $\alpha$  by demonstrating the existence of chimera 
in ensembles of identical Lorenz oscillators~\cite{enl1963}

\begin{align}
\dot{x}_{i}=&\,\sigma(y_{i}-x_{i})+\frac{\varepsilon}{2P}\sum_{j=i-P}^{j=i+P}~[\alpha_{11}(x_{j}-x_{i})+\alpha_{12}(y_{j}-y_{i})], \nonumber\\
\dot{y}_{i}=&\,x_{i}(\rho-z_{i})-y_{i}+\frac{\varepsilon}{2P}\sum_{j=i-P}^{j=i+P}~[\alpha_{21}(x_{j}-x_{i})+\alpha_{22}(y_{j}-y_{i})], \nonumber  \\
\dot{z}_{i}=&\,x_{i}y_{i}-b z_{i}, 
\label{eqlor}
\end{align}

and Hindmarsh-Rose (HR) neurons~\cite{msdl2008}

\begin{align}
\dot{x}_{i}=&\,y_{i}-ax_{i}^{3}+bx_{i}^{2}-z_{i}+I\\
&\,+\frac{\epsilon}{2P}\sum_{j=i-P}^{j=i+P}~[\alpha_{11}(x_{j}-x_{i})+\alpha_{12}(y_{j}-y_{i})], \nonumber \\
\dot{y}_{i}=&\,c-dx_{i}^{2}-y_{i}+\frac{\epsilon}{2P}\sum_{j=i-P}^{j=i+P}~[\alpha_{21}(x_{j}-x_{i})+\alpha_{22}(y_{j}-y_{i})],\nonumber  \\
\dot{z}_{i}=&\,r_{0}[s(x_{i}-x_{0})-z_{i}], 
\label{eqhr}
\end{align}

where $i=1,2,...,N=500$.  In Eq.~(\ref{eqlor}) $\sigma=10$, $\rho=28$, $b=8/3$ are the system parameters of
the individual Lorenz oscillators, while in Eq.~(\ref{eqhr})
$a=c=1$, $d=5$, $r_{0}=0.005$, $s=4$, $x_{0}=-1.6$, $b=3$, $I=3$ are the parameters of
the HR neurons. For the chosen parameter values, the individual Lorenz oscillators and HR neurons exhibit chaotic behavior.
The spatiotemporal plot and the average frequencies of the individual oscillators are 
shown in Fig.~\ref{fig6}(a) and (b), respectively, for $\alpha=1.46$, $r=0.35$ and $\varepsilon=3.5$
elucidating the existence of frequency chimera in the ensemble of Lorenz oscillators. 
It is to be noted that the frequency chimera is indeed a multichimera with four coherent and
incoherent domains as can be clearly visualized in both the spatiotemporal and the average frequency
plots. 
 To be more clear, the oscillators simply populate the two different
lobes of the attractor resulting in four coherent domains (see Fig. ~\ref{fig6}(a)),
as can be readily seen in the time averaged frequency $\omega_i$ in Fig. ~\ref{fig6}(b),
 separated by the incoherent borders due to non-local
coupling. Note that chimeras and multichimeras can also be further distinguished 
quantitatively, by introducing the measure of discontinuity as shown 
in~\cite{rgvkc2014}, though this is not pursued here. The strength of the 
incoherence $S$ for the amplitude and $S_{\omega}$ in the frequency domain  
for the ensemble of Lorenz oscillators are depicted in Fig.~\ref{fig6}(c) as a function
of the coupling strength $\varepsilon\in(0,5)$.  The null(unit) value of $S_{\omega}$($S$)
confirms that the frequencies(amplitudes) are synchronized(desynchronized) in the
range  $\varepsilon\in(0,1.9)$.  The fluctuation in the values of the strength of incoherence
$S_{\omega}$ and $S$ corroborates the existence of frequency chimera in the range of the 
coupling strength $\varepsilon\in(1.9,4.6)$.  Further, the inset of Fig.~\ref{fig6}(c) indicates an
increase in the randomness in the frequencies of the identical Lorenz oscillators for the values of
$\varepsilon$ where we observe the frequency chimera. Beyond $\varepsilon=4.6$, the zero values for
both $S_{\omega}$ and $S$ indicate the emergence of complete coherence among the
oscillators of the ensemble.

A two parameter bifurcation diagram as a function of the coupling strength $\varepsilon\in(0,5)$ 
and the asymmetry parameter  $\alpha\in(0,\frac{\pi}{2})$
of the ensemble of identical Lorenz oscillators is shown in Fig.~\ref{fig6a}. We have used the
same symbols as in Fig.~\ref{fig5} to indicate the different emergent behaviors. As 
in the case of the ensemble of R\"ossler oscillators, the ensemble of Lorenz oscillators
evolve asynchronously in the entire range of the coupling strength $\varepsilon\in(0,5)$ 
for $\alpha=\frac{\pi}{2}$. In the range of the asymmetry parameter  $\alpha\in(1.57,1.05)$, there
is a transition from asynchronization ($DSYC$) to frequency chimera ($FC$) and 
then to coherent states ($CO$) followed by complete synchronization ($CS$) as a 
function of the coupling strength. Above $\alpha=1.05$, the dynamical transition 
is in the sequence of desynchronization ($DSYC$), amplitude chimera ($AC$), 
cluster states ($CL$), coherent states ($CO$) and complete synchronization ($CS$)
upon increasing the coupling strength from null value.

\begin{figure}
\centering
\includegraphics[width=0.9\columnwidth]{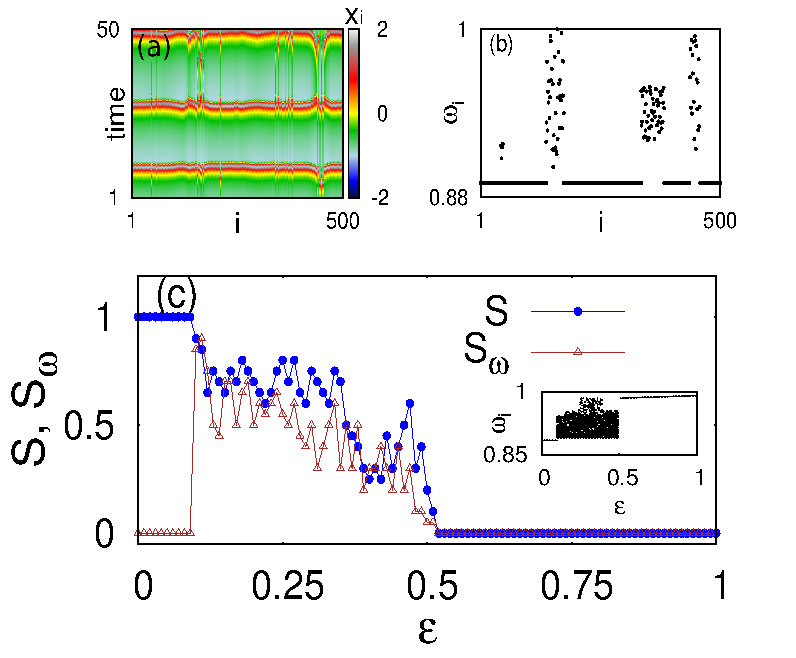}
\caption{(Color online)~(a) Space time plot and (b) time averaged frequency as a function of
the oscillator index for $\varepsilon=0.3$, and (c) $S$ and  $S_{\omega}$ as a function of the
coupling strength elucidating the frequency chimera in an ensemble of identical
Hindmarsh-Rose (HR) neurons with nonlocal rotational matrix coupling  for $r=0.35$ and $\alpha=1.46$.}
\label{fig7}
\end{figure}
\begin{figure}
\centering
\includegraphics[width=0.9\columnwidth]{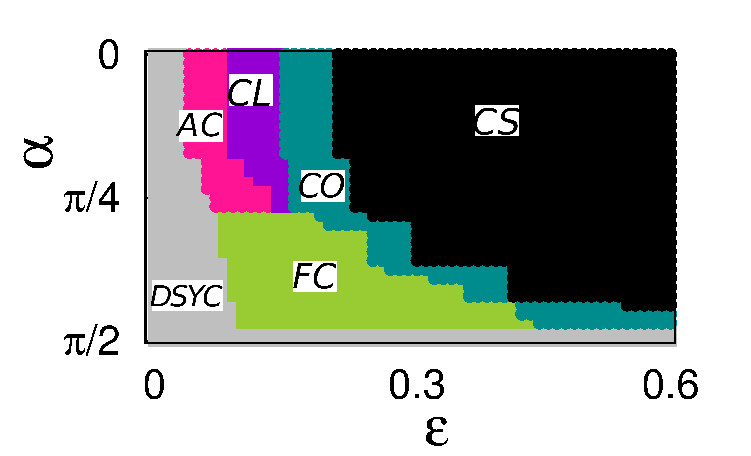}
\caption{(Color online)~Two parameter phase diagram for the dynamical states of the HR neurons
(\ref{eqhr}) as a function of the coupling strength
$\epsilon$ and the asymmetry parameter $\alpha$. See Table I and text for details.}
\label{fig7a}
\end{figure}

The existence of frequency chimera (multichimera), that is coexistence of coherence and incoherence
domains in the ensemble of HR neurons for $\alpha=1.46$, $r=0.35$ and $\varepsilon=0.3$
is shown in the spatiotemporal plot (see Fig.~\ref{fig7}(a)) and the 
average frequency plot (see Fig.~\ref{fig7}(b)). The dynamical transition from
asynchronous state to complete coherence via frequency chimera in the ensemble of HR neurons is evident
from the values of $S_{\omega}$ and $S$ in Fig.~\ref{fig7}(c). The coherent and incoherent
oscillations in frequency and amplitude, respectively, are revealed from the values of
 $S_{\omega}=0$ and $S=1$ in the range  $\varepsilon\in(0,0.09)$. Intermediate values of 
$S_{\omega}$ and $S$, that is between null value and unity, in the range of the coupling 
strength $\varepsilon\in(0.09,0.51)$ confirm the existence of frequency chimera.  As in
the ensemble of R\"ossler and Lorenz oscillators, the frequency chimera is accompanied by an
increase in the randomness in the frequencies of the ensemble of identical HR neurons.
The ensemble of HR neurons are in complete harmony for $\varepsilon>0.51$ as 
is clear from the values of $S_{\omega}$ and $S$ from Fig.~\ref{fig7}(c).
 A two parameter phase diagram as a function of the coupling strength $\varepsilon\in(0,0.6)$ 
and the asymmetry parameter  $\alpha\in(0,\frac{\pi}{2})$
of the ensemble of HR neurons is shown in Fig.~\ref{fig7a}. We have followed
the same abbreviations as in Figs.~\ref{fig5} and \ref{fig6a}  to indicate the different dynamical regimes here also.  All the HR neurons evolve independently for
$\alpha=\frac{\pi}{2}$ in the entire range of the $\varepsilon$ we have analyzed.
There is a transition from asynchronous state to synchronous state via frequency chimera
followed by coherent states as a function of the coupling strength for $\alpha\in(1.57,0.86)$.
For further larger values of the asymmetry parameter, the frequency chimera is replaced  by the 
amplitude chimera and cluster states in the above dynamical transition as seen in Fig.~\ref{fig7a}.

\section{Conclusion}
In conclusion, we have shown that the nonlocal rotational matrix coupling can indeed 
induce chimera states in an ensemble of a generic nonlinear dynamical systems 
exhibiting chaotic oscillations that cannot be
reduced to phase models, whereas dynamical systems in the periodic regimes facilitating
phase reduction have been employed so far along with such coupling.
In particular, using an ensemble of identical
R\"ossler systems, exhibiting chaotic oscillations for appropriate parameter
values, we have demonstrated that frequency and amplitude chimeras are induced
by the nonlocal rotational matrix coupling in a rather large range of 
the asymmetry parameter $\alpha$ as a function of the coupling strength in contrast
to the earlier studies limiting to near $\frac{\pi}{2}$ values of it
in the phase-only models and in the dynamical systems that can be phase reduced.
Further the chimera states
are characterized by inhomogeneous spread of coherent and incoherent domains
unlike homogeneous spread of these domains in phase models
and in periodic oscillators. 
In addition, a range of collective behaviors including cluster states, coherent states and
complete synchronization are observed during the dynamical transitions as a function
of the system parameters. We have also corroborated the emergence of frequency
and amplitude chimeras using the strength of incoherence in the frequency domain ($S_{\omega}$)
and in the amplitude domain ($S$). The generic nature of the nonlocal rotational matrix coupling
in inducing chimera in any nonlinear dynamical systems in a wide range of 
the asymmetry parameter is also confirmed by illustrating the emergence of chimeras 
in an ensemble of identical Lorenz systems and Hindmarsh-Rose neurons in their chaotic regimes.
 Further, increase in the randomness of the frequencies of the ensemble of identical oscillators
may be a signature  of the frequency chimera as observed in all the three paradigmatic models
we have analyzed.
Thus it is clear that the asymmetry parameter in the nonlocal rotational matrix coupling,
which induces the asymmetry in the ensemble of identical nonlinear systems,
plays a crucial role in determining the desired dynamical behavior. 
Hence for appropriate choice of the asymmetry parameter, it is possible to obtain the
desired collective states in dynamical systems including neuronal systems~\cite{kozma1998}, 
hydrodynamical systems such as laminar and turbulent regions in Couette flow studies~\cite{barkley2012,brethouwer2012}, etc.,
with appropriate nonlocal coupling.

\section*{Acknowledgments}
%\begin{acknowledgments}
The work of VKC is supported by INSA young scientist project.
DVS is supported by the SERB-DST Fast Track scheme for young
scientist under Grant No. ST/FTP/PS-119/2013.
The work of RG and ML is supported by a Department of Science and Technology (DST), Government
of India, IRHPA research project. ML is also supported by a DAE Raja Ramanna
Fellowship.
%\end{acknowledgments}

\end{document}